\documentclass[doublecol,linenumbers]{epl2}

\usepackage{graphicx}
\usepackage{subfigure}
\usepackage{amsthm}
\usepackage{amsmath}
\usepackage{amssymb}
\usepackage{verbatim}
\usepackage{dcolumn}
\usepackage{bm}
\usepackage{epsf}
\usepackage{color}
\usepackage[colorlinks=true,citecolor=blue,linkcolor=blue,urlcolor=blue]{hyperref}%
\usepackage{xcolor}
\usepackage{dsfont}
\newcommand{\bra}[1]{\left\langle #1\right|}
\newcommand{\ket}[1]{\left|#1\right\rangle}
\newcommand{\braket}[2]{\left\langle #1|#2\right\rangle}

\newcommand{\tr}[1]{\mathrm{tr}\left\{#1\right\}}

\newcommand{\la}{\left\langle}
\newcommand{\ra}{\right\rangle}
\newcommand{\pd}{\partial}

\newcommand{\td}{\mathrm{d}}

\newcommand{\e}[1]{\exp{\left(#1\right)}}
\newcommand{\lo}[1]{\ln{\left(#1\right)}}

\newcommand{\com}[2]{\left[#1,\,#2\right]}

\newcommand{\bla}{bla\\bla\\bla\\bla\\bla}
\newcommand{\mb}[1]{\mbox{\boldmath$#1$}}
\newcommand{\mc}[1]{\mathcal{#1}}
\newcommand{\mbb}[1]{\mathbb{#1}}
\newcommand{\mbf}[1]{\mathbf{#1}}
\newcommand{\mf}[1]{\mathfrak{#1}}
\newcommand{\mrm}[1]{\mathrm{#1}}

\DeclareMathOperator*{\sumint}{%
\mathchoice%
  {\ooalign{$\displaystyle\sum$\cr\hidewidth$\displaystyle\int$\hidewidth\cr}}
  {\ooalign{\raisebox{.14\height}{\scalebox{.7}{$\textstyle\sum$}}\cr\hidewidth$\textstyle\int$\hidewidth\cr}}
  {\ooalign{\raisebox{.2\height}{\scalebox{.6}{$\scriptstyle\sum$}}\cr$\scriptstyle\int$\cr}}
  {\ooalign{\raisebox{.2\height}{\scalebox{.6}{$\scriptstyle\sum$}}\cr$\scriptstyle\int$\cr}}
}

\title{Thermodynamic control -- an old paradigm with new applications}

\author{Sebastian Deffner \inst{1} \and Marcus V. S. Bonan\c{c}a\inst{2}}
\shortauthor{Sebastian Deffner and Marcus V. S. Bonan\c{c}a}

\institute{     \inst{1} Department of Physics, University of Maryland, Baltimore County, Baltimore, MD 21250, USA\\
  \inst{2} Instituto de F\'isica `Gleb Wataghin', Universidade Estadual de Campinas, 13083-859, Campinas, S\~{a}o Paulo, Brazil
}

\abstract{Tremendous research efforts have been invested in exploring and designing so-called shortcuts to adiabaticity. These are finite-time processes that produce the same final states that would result from infinitely slow driving. Most of these techniques rely on auxiliary fields and quantum control techniques, which makes them rather costly to implement. In this Perspective we outline an alternative paradigm for optimal control that has proven powerful in a wide variety of situations ranging from heat engines over chemical reactions to quantum dynamics -- thermodynamic control. Focusing on only a few, selected milestones we seek to provide a pedagogical entry point into this powerful and versatile framework.}

\begin{document}

\maketitle

\section{Introduction}

The desire to control our environment and all processes happening therein appears to be deeply rooted in human nature \cite{Leotti2010}. More mundanely, it is a ubiquitous goal in physics and engineering to identify optimal processes that waste the minimal amount of resources to achieve a predetermined goal. In essence, this is also the core motivation of thermodynamics that was originally designed to understand and optimize heat engines \cite{Kondepudi1998}. Since, however, thermodynamics in its traditional formulation is restricted to idealized situations and infinitely slow processes \cite{Callen1985}, the development of extensions and generalizations to finite-time processes was inevitable.

A particularly fruitful approach is based on expressing the nonequilibrium entropy production, which sometimes is also called dissipated availability or excess work, as a geometric form in terms of the thermodynamic metric \cite{Salamon1983,Ruppeiner1995}. Despite its promising beginnings almost four decades ago, the field remained somewhat unrecognized by physicists until very recently the deep connection of thermodynamic control and stochastic thermodynamics was unveiled \cite{crooks2007,schmiedl2007,aurell2011,Sivak2012,bonanca2014}. Since then thermodynamic control strategies have slowly but steadily been attracting more attention, since they provide universal means to suppress nonequilibrium excitations in classical \cite{sivak2016,Bonanca2016,large2018,Bonanca2018,bonanca2019} and quantum systems \cite{Acconcia2015,zulkowski2015b,Mancino2018,scandi2019}. 

Concurrently, yet independently quantum control experienced a surge of development with the discovery of so-called shortcuts to adiabaticity (STA) \cite{STAreview}. A STA is a finite-time process with the same final state that would result from infinitely slow, adiabatic driving. Over the last decade, STAs have developed into a tremendous field of modern research, whose different methods and techniques are comprehensively reviewed in ref.~\cite{STAreview}. Thus, it is rather surprising that despite extensions of STA to classical systems \cite{deffner2014PRX,Le2016,Patra2017,Iram2019}, thermodynamic control of quantum dynamics has not gained all that much attention, yet.

Actually, it is not a big leap to recognize that thermodynamic control strategies are uniquely suited to facilitate STA in, e.g., quantum annealing \cite{Johnson2011}, where most other techniques of STA are neither practical nor realistic \cite{Deffner2019book}. If a system is originally prepared in its ground state, then any excitations will necessarily be accompanied by nonequilibrium work (or entropy production). Suppressing exactly this nonequilibrium work is what thermodynamic control strategies are designed to achieve.

In the context of thermodynamic control, STA can be phrased in a wider sense as means to obtain finite-time protocols that yield results usually achievable only by quasistatic processes. The second law of thermodynamics asserts that driving a system from a given thermodynamic state (in contrast to a quantum state) to another is always accompanied by a cost, that is typically minimal for quasistatic or reversible driving. A \emph{finite-time process} that performs the same task at minimal or even zero cost is then a \emph{thermodynamic shortcut} \cite{martinez2016,li2017,dann2019}.

In this Perspective, we outline some of the hallmarks of the substantial amount of work that has been accumulated on thermodynamic control. Our goal is to provide a concise and introductory overview of the beginnings and the recent developments in the hope that thermodynamic control will experience wider application as STA. 

\section{Minimal dissipation and thermodynamic length}

We start at the beginning and the earliest accounts. Aiming to determine limits on the efficiency of finite-time processes in thermodynamic systems \cite{andresen1977,salamon1977a,salamon1977b,salamon1980,salamon1982}, Salamon and Berry initiated the study of thermodynamic control \cite{Salamon1983}. To this end, they focused on  \emph{endoreversible} processes \cite{Curzon1975,Rubin1979}.  Such processes \cite{Hoffmann1997} are slow enough for the system  to \emph{locally equilibrate}, yet the processes are too fast for the system to reach a state of equilibrium with the environment. Thus, at any instant the internal energy $U(\mbf{X})$ is well-defined, where $\mbf{X}=X_1,\dots,X_n$ are extensive parameters.

For small displacements away from equilibrium, $\Delta \mbf{X}=\mbf{X}-\mbf{X}^\mrm{eq}$, we can expand $U$ in powers of $\Delta \mbf{X}$,
\begin{equation}
\label{eq:deltaU}
\Delta U=\frac{1}{2}\sum_{i,j} \eta_{ij}\, \Delta X_i\,\Delta X_j
\end{equation}
where $\eta_{ij}$ is the  thermodynamic metric \cite{Weinhold1975,Ruppeiner1995},
\begin{equation}
\label{eq:eta}
\eta_{ij}=\frac{\pd^2 U}{\pd X_i\,\pd X_j}\,.
\end{equation}
Note that the linear term in eq.~\eqref{eq:deltaU} vanishes, since $U(\mbf{X})$ is minimal in equilibrium \cite{Callen1985}. Correspondingly, we have for the intensive parameters $Y_i=\pd U/\pd X_i$,
\begin{equation}
\Delta Y_i=\sum_j \eta_{ij}\,\Delta X_i\,,
\end{equation}
which is an expression of Le Chatelier's principle \cite{Callen1985}.

Now consider an endoreversible process that is driven by a slow variation of the extensive parameters, $\mbf{X}(t)$ for $0\leq t \leq\tau$. Denoting the intensive parameters of the environment by $\mbf{Y}^\mrm{e}(t)$, the dissipated availability or exergy \cite{Schlogl1989} reads,
\begin{equation}
\Delta \mc{E}=\int_0^\tau dt\,\sum_i \left(Y_i(t)-Y_i^\mrm{e}(t)\right)\, \dot{X}_i\,,
\end{equation}
where we denote a derivative with respect to time by a dot. The intensive parameters of the environment can then be determined by,
\begin{equation}
\mbf{Y}^\mrm{e}(t)-\mbf{Y}(t)=\eta\big|_{\mbf{X}(t)}\,\left(\mbf{X}^\mrm{e}(t)-\mbf{X}(t)\right)\,,
\end{equation}
which is well-defined since $\eta$ is non-singular.

It is  a simple exercise to show from geometric considerations that we have \cite{Salamon1983}
\begin{equation}
\label{eq:exergy}
\Delta \mc{E}=\bar{\epsilon} \int_0^\tau dt\, \dot{\mbf{X}}^\top\, \eta\, \dot{\mbf{X}}\,,
\end{equation}
where $\bar{\epsilon}$ is the average \emph{lag} time given by, $\mbf{X}^\mrm{e}(t)\simeq \mbf{X}(t)+\epsilon\,\dot{\mbf{X}}$. Note that the average is taken over time, and hence the lag time depends on the parameterization.

Applying the Cauchy-Schwarz inequality to eq.~\eqref{eq:exergy}, we can write
\begin{equation}
\label{eq:thermo_length}
\Delta \mc{E}\geq \frac{\bar{\epsilon}}{\tau}\, \left[\int_0^\tau dt\, \sqrt{\dot{\mbf{X}}^\top\, \eta\, \dot{\mbf{X}}}\right]^2\equiv \frac{\bar{\epsilon}\,\ell^2}{\tau}  \,,
\end{equation}
where we identified the thermodynamic length $\ell$ \cite{Weinhold1975}. An immediate consequence of eq.~\eqref{eq:thermo_length} is that processes with given mean lag time, $\bar{\epsilon}$, dissipate the minimal availability when the system operates at constant speed. In other words, thermodynamically optimal processes are given by the geodesics under the thermodynamic metric tensor $\eta$.

In the years following the original account \cite{Salamon1983}, thermodynamic control did gain some attention \cite{andresen1984,Nulton1985,Schlogl1985,Spirkl1995}, in particular generalized to the entropy representation of thermodynamics \cite{Andresen1984CR,Salamon1984JCP,Salamon1985,Feldmann1986}. However, until its rediscovery by Crooks \cite{crooks2007} the area of research had escaped the broad attention of the physics community \cite{Schon1993,Andresen1996,andresen2011}.

Whereas Salamon's and Berry's approach \cite{Salamon1983} is formulated in terms of thermodynamic notions, Crooks analysis \cite{crooks2007} is rooted in statistical mechanics. Consider a physical system in equilibrium with a thermal environment. Then the equilibrium Gibbs distribution with partition function $Z$ can be written as
\begin{equation}
\label{eq:dist}
p(\Gamma,\lambda)=\frac{1}{Z}\,\e{\sum_i \lambda_i(t) X_i(\Gamma)}\,,
\end{equation}
where $X_i(\Gamma)$ is an extensive variable evaluated  at a point in phase space $\Gamma$, and $\lambda_i(t)$ are the generalized conjugate forces. The logarithm of the partition function is related to the Massieu potential $\Psi$ \cite{Callen1985}, and we can write
\begin{equation}
\lo{Z}=\Psi=S-\sum_i \lambda_i \la X_i\ra\,.
\end{equation}
Note that the $\Psi$ is nothing else but the Legendre transform of the thermodynamic entropy $S$ \cite{Callen1985}. Its covariance matrix,
\begin{equation}
\label{eq:g}
g_{ij}=\frac{\pd^2 \Psi}{\pd \lambda_i  \pd \lambda_j}\,,
\end{equation}
defines a Riemannian metric. Comparing eq.~\eqref{eq:eta} with eq.~\eqref{eq:g} we recognize $g$ as the entropy representation of the thermodynamic metric, $\eta$. 

However, combining eq.~\eqref{eq:g} with eq.~\eqref{eq:dist} we also immediately identify $g$ as the Fisher information matrix of the instantaneous equilibrium distribution $p(\Gamma,\lambda)$,
\begin{equation}
g_{ij}(\lambda)=\sumint_\Gamma\,p(\Gamma) \frac{\pd \lo{p(\Gamma,\lambda)}}{\pd \lambda_i}\frac{\pd \lo{p(\Gamma,\lambda)}}{\pd \lambda_j}\,.
\end{equation}
Thus, the thermodynamic length \eqref{eq:thermo_length} can be written as
\begin{equation}
\ell=\int_0^\tau dt \sqrt{\sumint_\Gamma \frac{\left(\dot{p}(\Gamma,\lambda)\right)^2}{p(\Gamma,\lambda)} }\,,
\end{equation}
which is nothing else but Wootters' statistical distance \cite{Wootters1981} measuring the distinguishability of the distributions $p(\Gamma,\lambda(0)$ and $p(\Gamma,\lambda(\tau))$. Finally, introducing the thermodynamic divergence, $\mc{J}\equiv \tau \Delta \mc{E}/\bar{\epsilon}$, Crooks found \cite{crooks2007},
\begin{equation}
\label{eq:crooks}
\mc{J}\geq \ell^2\,,
\end{equation}
which is equivalent to Salamon's and Berry's finding \eqref{eq:thermo_length}.

Crooks' contribution is important for two reasons: (i) his analysis highlighted the close connection of thermodynamically optimal processes and information geometry, which is not necessarily restricted to endoreversible processes. This generality was already hinted at by Salamon and Berry \cite{Salamon1983}, yet it had not been discussed this transparently before. And (ii), the mathematical tools employed in ref.~\cite{crooks2007} lend themselves naturally to generalize thermodynamic control by means of linear response theory and to extend the approach to genuinely quantum dynamics.

\section{Thermodynamic control from linear response}

Both, Salamon's and Berry's approach \cite{Salamon1983} as well as Crooks' account \cite{crooks2007}, are explicitly built on the notion of endoreversibility. To generalize the geometric framework for thermodynamic control to a broader class of situations, Sivak and Crooks \cite{Sivak2012} re-derived the thermodynamic length analysis directly from linear response theory \cite{kubo1985}. As main results \cite{Sivak2012}, it was shown that optimal driving protocols with minimal dissipation are geodesics on the thermodynamic manifold, that dissipation is inversely proportional to process duration, that the optimal control protocol is independent of duration, and that optimal protocols are characterized by constant power.
\begin{figure}
\centering
\includegraphics[width=.4\textwidth]{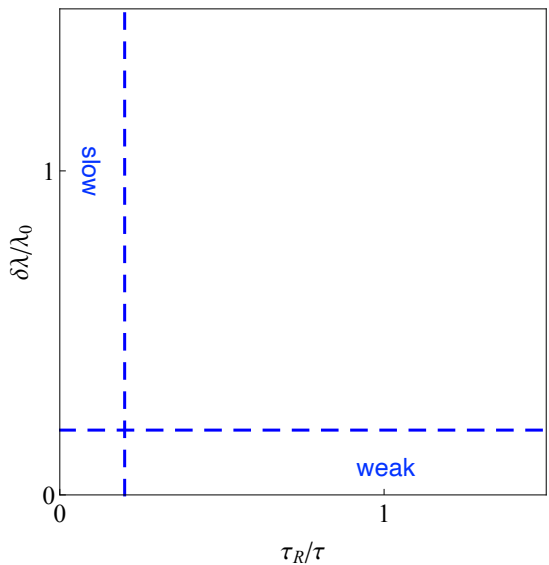}
\caption{\label{fig:regime} Illustration of the dynamical parameter space. Linear response frameworks for thermodynamic control have been developed for slow, $\tau_R/\tau \ll 1$, and for weak, $\delta \lambda/\lambda_{0} \ll 1$ driving.}
\end{figure}

\subsection{Slowly varying processes -- endoreversibility 2.0}

More generally, linear response methods can be employed for systems that are only weakly driven, and for systems that remain close to equilibrium at all instants, cf. fig.~\ref{fig:regime}. Sivak's and Crooks' approach \cite{Sivak2012} falls under so-called \emph{slowly varying processes}. In this scenario, a system is driven slowly enough such that it quickly returns to a state of equilibrium. In formula we have, $\tau_R/\tau\ll 1$, where $\tau_R$ is the relaxation time.  Thus, slowly varying processes can be considered a generalization of endoreversible processes during which a system \emph{remains} in local equilibrium.

For the sake of consistency and simplicity, we will be continuing the discussion in our notation for a single control parameter $\lambda(t)$ \cite{bonanca2014} that is varied from $\lambda(0)=\lambda_0$ to $\lambda(\tau)=\lambda_0+\delta\lambda$ during time $\tau$. For infinitely slow variation, i.e., in the limit $\tau\rightarrow \infty$ the work performed by the system is given by the free energy difference, $\Delta F \equiv F(\beta;\lambda_0+\delta\lambda) - F(\beta;\lambda_{0})$, where as always, $F(\beta;\lambda)=-1/\beta\,\ln Z(\beta,\lambda)$ and $\beta$ is the inverse temperature. The maximum work theorem, now predicts that for all finite values of $\tau$ we have $\la W_\mrm{ex}\ra=\la W\ra -\Delta F\geq 0$, which means that for all realistic, irreversible processes excess work $\la W_\mrm{ex}\ra$ is dissipated into the environment. From stochastic thermodynamics \cite{jarzynski_1997_PRE} we have
\begin{equation}
\label{eq:work}
\la W \ra = \int_{0}^{\tau} dt\,\dot{\lambda} \,\la \frac{\partial H}{\partial\lambda}\ra\,.
\end{equation}
where the angular brackets denote an average over many realizations of the process.

Equation~\eqref{eq:work} is the starting point for any treatment of thermodynamic control with linear response theory. Discretizing in time and expanding the Hamiltonian in each interval in linear order of $\delta\lambda$ it can be shown \cite{Sivak2012,bonanca2014} that
\begin{equation}
\label{eq:work_slow}
\la W_{\rm{ex}}\ra = \beta\int_{0}^{\tau}\td t\,\dot{\lambda}^{2}(t) \tau^{c}[\lambda(t)]\,\mathcal{X}[\lambda(t)]\,,
\end{equation}
where $\mathcal{X}[\lambda(t)]$ is the variance of the observable
\begin{equation}
\mathcal{X}[\lambda(t)] = \left\langle \left( \frac{\partial H}{\partial\lambda}\right)^{2} \right\rangle_{\lambda(t)} - \left\langle \frac{\partial H}{\partial\lambda}\right\rangle_{\lambda(t)}^{2}\,.
\end{equation}
Note, that  $ \mathcal{X}[\lambda(t)]$  is closely related to the Fisher information for $\pd H/\pd\lambda$ under nonequilibrium dynamics \cite{frieden2000}. Moreover, $\tau^{c}[\lambda(t)]$ is the \emph{correlation time}, which is determined by the relaxation function of the system \cite{bonanca2014}. 

For the purposes of control, we immediately observe that eq.~\eqref{eq:work_slow} expresses $\la W_{\mathrm{ex}}\ra$ as a functional of $\lambda(t)$. Numerically this functional \eqref{eq:work_slow} was studied previously by de Koning \cite{dekoning2005}, where, however, the correlation time, $\tau^{c}(\lambda) $,  and the variance, $\mathcal{X}(\lambda) $, were only obtained numerically. Generally, it is rather straight forward to determine analytical expressions for $\mathcal{X}(\lambda) $, whereas treating the correlation time is more involved. We showed in ref.~\cite{bonanca2014} that also the correlation time can be obtained systematically from microscopic properties.

This framework has proven powerful  to find optimal driving protocols for a wide range of problems, including, e.g., molecular machines and biological applications \cite{zulkowski2012,zulkowski2015a,sivak2016,large2018,lucero2019,large2019,blaber2020}, many body systems \cite{rotskoff2015,gingrich2016,rotskoff2017}, nonequilibrium phase transitions \cite{deffner2017},  optimal performance of heat engines \cite{bonanca2019,brandner2020,Abiuso2020}, transitions between nonequilibrium stationary states \cite{zulkowski2013,mandal2016}, and bit erasure \cite{zulkowski2014}.

\subsection{Slowly varying quantum processes}

Shortly after the framework of slowly varying process gained some prominence, it was also generalized to quantum dynamics by Zulkowski and DeWeese \cite{zulkowski2015b}. They considered a joint quantum system that evolves unitarily under
\begin{equation}
H_\mrm{tot}=H_\mrm{sys}(t)+H_B+\gamma \sum_\alpha A_\alpha\otimes B_\alpha\,,
\end{equation}
where $H_\mrm{syts}(t)$ and $H_B$ are the reduced Hamiltonians of system and bath respectively. The (weak) interaction terms are composed of Hermitian operators $A_\alpha$ acting on the system, and $B_\alpha$ acting on the bath. 

In the ultraweak coupling approximation, $\gamma\ll 1$, and under the usual assumptions, the reduced dynamics for the system only becomes a Lindblad master equation \cite{Gorini1976},
\begin{equation}
\label{eq:lind}
\dot{\rho}=-\frac{i}{\hbar}\com{H(t)}{\rho(t)}+\gamma^2\,\mc{D}(\rho(t))\,,
\end{equation}
where the dissipative part reads
\begin{equation}
\begin{split}
\mc{D}(\rho(t))&=\sum_{\alpha,\beta}\sum_\omega\,\kappa_{\alpha\beta}(\omega)\,\Big(L_{\omega,\beta}(t)\rho(t)L^\dagger_{\omega,\alpha}(t)\\
&\quad-\frac{1}{2}\left\{L^\dagger_{\omega,\alpha}(t)L_{\omega,\beta}(t),\rho(t)\right\}\Big)\,.
\end{split}
\end{equation}
Note that $H(t)=H_\mrm{sys}(t)+\gamma^2 H_{LS}(t)$, where $H_{LS}(t)$ is the Lamb-shift due to the coupling of the system with the thermal bath.

It has been shown \cite{Deffner2011} that the irreversible entropy production rate, i.e., the rate with which excess work is produced, can be written as,
\begin{equation}
\sigma_\mrm{ex}(t)= -\tr{\dot{\rho}\left(\lo{\rho(t)}-\lo{\rho^\mrm{eq}(t)}\right)}\,,
\end{equation}
where $\rho^\mrm{eq}(t)=\e{-\beta H_\mrm{sys}(t)}/Z$ is the \emph{instantaneous} Gibbs state. Writing the excess work as, $\beta \la W_\mrm{ex}\ra=\int_0^\tau dt\, \sigma_\mrm{ex}(t)$, a quantum version of eq.~\eqref{eq:work_slow} can be found.

In complete analogy to the classical case, Zulkowski and DeWeese  \cite{zulkowski2015b} assumed that the driven process varies only slightly from endoreversibility. Thus, they wrote
\begin{equation}
\rho(t)=\rho^\mrm{eq}(t)+\sum_\alpha \delta\rho_\alpha\,\dot{\lambda}_\alpha\,,
\end{equation}
where as before the $\lambda_\alpha$ denote conjugate forces. It can then be shown by tedious, but straight forward math that
\begin{equation}
\begin{split}
&\beta \la W_\mrm{ex}\ra=\\
&\quad\gamma^2\int_0^\tau dt \sum_{jklm}\sum_{\alpha,\beta} \dot{\lambda}_\alpha \dot{\lambda}_\beta \mc{A}_{jklm}\left(\pd_{\lambda_\alpha}\rho^\mrm{eq}\right)_{jk}\left(\pd_{\lambda_\beta}\rho^\mrm{eq}\right)_{lm}\,.
\end{split}
\end{equation}
The tensor $ \mc{A}_{jklm}$ has a rather involved expression, which can be found in the appendix of ref.~\cite{zulkowski2015b}.

Similarly to before, the excess work can be written as a quadratic form. Thus, also in quantum dynamics optimal driving protocols can be determined relatively easily by standard means of variational calculus. However, the physical interpretation of $\mc{A}$ is somewhat obscured by its mathematical complexity. Therefore, an alternative approach that more closely resembles the thermodynamic length \eqref{eq:thermo_length} appears desirable.

\section{Quantum thermodynamic length}

Efforts to define the quantum thermodynamic length had been undertaken already by Deffner and Lutz \cite{deffner2010}. To this end, eq.~\eqref{eq:thermo_length} needed to be generalized to an expression for the quantum entropy production, first. For isolated quantum systems that are initially prepared in thermal equilibrium, the excess work is given by the relative entropy \cite{deffner2010}
\begin{equation}
\label{eq:rel}
\beta \la W_\mrm{ex}\ra=\tr{\rho_\tau\lo{\rho_\tau}}-\tr{\rho_\tau\lo{\rho^\mrm{eq}_\tau}}=S(\rho_\tau||\rho_\tau^\mrm{eq})\,,
\end{equation}
where $\rho_\tau$ is the quantum state at the end of the process and $\rho_\tau^\mrm{eq}$ is the Gibbs state corresponding to the final Hamiltonian. Expanding the relative entropy the first non-vanishing order can be identified as the squared Bures angle between $\rho_\tau$ and $\rho_\tau^\mrm{eq}$ \cite{deffner2010,deffner2013},
\begin{equation}
\label{eq:qm_length}
\beta \la W_\mrm{ex}\ra\geq \frac{8}{\pi^2}\, \mc{L}^2(\rho_\tau,\rho_\tau^\mrm{eq})\,,
\end{equation}
where the Bures angles is defined in terms of the quantum fidelity, $F(\rho_\tau,\rho_\tau^\mrm{eq})=\left[\tr{\sqrt{\sqrt{\rho_\tau}\rho_\tau^\mrm{eq}\sqrt{\rho_\tau}}}\right]^2 $ as
\begin{equation}
\mc{L}(\rho_\tau,\rho_\tau^\mrm{eq})=\arccos\sqrt{F(\rho_\tau,\rho_\tau^\mrm{eq})}\,.
\end{equation}

Comparing eq.~\eqref{eq:qm_length} with eqs.~\eqref{eq:thermo_length} and \eqref{eq:crooks}, we recognize $\mc{L}(\rho_\tau,\rho_\tau^\mrm{eq})$ as the quantum version of the thermodynamic length. The natural question arises whether this insight can be further specified and exploited in thermodynamic quantum control. The affirmative answer was given by Scandi and Perarnau-Llobet \cite{scandi2019} only very recently.

In complete analogy to the classical case \cite{Sivak2012,bonanca2014}, Scandi and Perarnau-Llobet \cite{scandi2019} considered an open quantum system undergoing slow driving. To this end, the dynamics is again described by a Lindblad master equation \eqref{eq:lind}, which we now write as $\dot{\rho}(t)=\mf{L} \left[\rho(t)\right]$, and whose stationary solution is the thermal Gibbs state, $\mf{L}\left[\rho^\mrm{eq}\right]=0$. Again expanding the Hamiltonian in terms of extensive observables, $X_i$, and conjugate forces, $\lambda_i$, such that $H=\sum \lambda_i X_i$, Scandi and Perarnau-Llobet  showed \cite{scandi2019}
\begin{equation}
\label{eq:work_qm}
\la W_\mrm{ex}\ra =\beta \int_0^\tau dt\, \sum_{i,j}\dot{\lambda}_i(t)\,\eta_{ij}^\mrm{qm}\,  \dot{\lambda}_j(t)\,,
\end{equation}
where the metric tensor now reads
\begin{equation}
\eta_{ij}^\mrm{qm} =-\frac{1}{2}\tr{X_i\, \mf{L}^D [\mbb{J}(X_j)]+X_j\, \mf{L}^D [\mbb{J}(X_i)]}\,.
\end{equation}
Here, $\mf{L}^D$ is the Drazin inverse of the quantum Liouvillian \eqref{eq:lind}, which first appeared in the treatment of controlling open classical systems \cite{mandal2016}. It reads for arbitrary $A$,
\begin{equation}
\mf{L}^D[A]=\int_0^\infty d\nu\,\e{\nu \mf{L}}\left(\rho^\mrm{eq}(t)\,\tr{A}-A\right)\,.
\end{equation}
Moreover, the operator $\mbb{J}$ is defined by
\begin{equation}
\mbb{J}[A]=\int_0^1 ds\,\rho^{1-s}\left(A-\tr{\rho A}\mbb{I}\right)\rho^s\,.
\end{equation}

Equation~\eqref{eq:work_qm} is the generalization of eqs.~\eqref{eq:thermo_length} and \eqref{eq:work_slow} to open quantum dynamics. However, due to the mathematical complexity of open quantum dynamics, the thermodynamic metric $\eta_{ij}^\mrm{qm}$ is no longer simply given by the Fisher information \eqref{eq:work_slow}. Rather remarkably, the derivation is only based on expressing the excess work as a relative entropy \cite{deffner2010,Deffner2011}, and assuming that the quantum system remains close to equilibrium at all times.

\section{Weak driving -- a blueprint for shortcuts to adiabaticity from thermodynamic control}

Despite the great success of the previous approach in elucidating the physics of optimal processes, for applications in, e.g., quantum annealing slowly-varying processes are not an adequate paradigm. The derivation of eqs.~\eqref{eq:work_slow} and \eqref{eq:work_qm} crucially depends on the assumption that the system rapidly equilibrates with a thermal environment. For a large class of isolated quantum dynamics, the treatment is thus not valid. On the other hand, linear response for weak driving, $\delta\lambda/\lambda_0 \ll 1$ (see fig.~\ref{fig:regime}), is uniquely suited to find optimal driving protocols with (approximately) suppressed nonequilibrium excitations. To this end, it proved useful to generalize the previous treatment of the excess work, $\la W_\mrm{ex}\ra$, \cite{Acconcia2015,acconcia2015a,pierre2020} to this new regime.

In this case, the central object is the so-called relaxation function $\Psi(t)$ whose expression is determined by the response function $\phi(t)$, via $\phi(t)=-\dot{\Psi}(t)$, with \cite{kubo1985}
\begin{equation}
\label{eq:response}
\phi(t)= \dfrac{1}{i\hbar} \,\tr{\rho^\mrm{eq}_0\,\com{\pd_\lambda H(0)}{\pd_\lambda H(t)}},
\end{equation}  
where we denote the generalized force as $\pd_\lambda H = \partial H/\partial\lambda$, and as before $\rho^\mrm{eq}_0=\e{-\beta H(\lambda_0)}/Z$. Consequently, the excess work \eqref{eq:work} can then be written as \cite{Acconcia2015} 
\begin{equation}
\label{eq:work_weak}
\la W_\mrm{ex}\ra= \frac{1}{2}  \int_{0}^{\tau} d t \int_{0}^{\tau} d s \, \dot{\lambda}(t)\, \Psi(t-s)  \, \dot{\lambda}(s)\,,
\end{equation}
which is again a quadratic form of the driven protocol $\lambda(t)$. Note, that the derivation of eq.~\eqref{eq:work_weak} does not require to assume any actual relaxation to occur. Rather, the excess work is full characterized by the response function \eqref{eq:response} of the quantum system. This means, in particular, that this framework also applies to unitary dynamics of isolated quantum systems. 

In the adiabatic limit no transitions between eigenstates occur, and therefore the excess work $\la W_\mrm{ex}\ra$ vanishes for infinitely slow driving. However, we showed in ref.~\cite{Acconcia2015} that optimal driving protocols exits, for which $\la W_\mrm{ex}\ra$ vanishes even in finite-time processes. These special driving protocols then constitute a STA. In ref.~\cite{Acconcia2015} we analyzed the parametric harmonic oscillator and driven qubits, of which we will briefly recount the latter. In addition, we showed in ref.~\cite{Bonanca2018} that the optimal protocols obtained from the quadratic form \eqref{eq:work_weak} closely resemble the optimal protocols for fully nonequilibrium dynamics -- more so than their slowly-varying analogs.

\subsection{Shortcut to adiabaticity for qubits}

Consider a qubit that is driven by  a time-dependent magnetic field subjected to the constraint $|\mathbf{B}(t)|=B_{0}=\mathrm{constant}$. Then, the  Hamiltonian reads 
\begin{equation}
H(t) = -\frac{\hbar\omega}{2}\,\boldsymbol\sigma\cdot\mathbf{B}(t)\,,
\label{spinhamil}
\end{equation}
where $\boldsymbol\sigma$ denotes the Pauli matrices. The magnetic field, $\mathbf{B}(t)$,  is parameterized by
\begin{equation}
\mathbf{B}(t) = B_{0}\,\begin{pmatrix}
\sin{[\varphi(t)]} \cos{[\theta(t)]}\\ \sin{[\varphi(t)]} \sin{[\theta(t)]}\\ \cos{[\varphi(t)]}
\end{pmatrix}\,.
\label{magfield}
\end{equation} 
and we write $\varphi(t) = \varphi_{0} + \delta\varphi\,g_{\varphi}(t)$ and $ \theta(t) = \theta_{0} + \delta\theta\,g_{\theta}(t)$ with the boundary conditions $g_{\varphi, \theta}(0) = 0$ and $g_{\varphi, \theta}(\tau) = 1$.

It is then easy to see \cite{Acconcia2015} that the excess work \eqref{eq:work_weak} is independent of $\theta(t)$. Hence, the response function \eqref{eq:response} can be written as
\begin{equation}
\phi(t) = \frac{\hbar}{2}\left(\omega B_{0}\right)^{2} \tanh{\left( \frac{\beta \hbar \omega B_{0}}{2}\right)} \sin{\left(\omega B_{0} t\right)}\,,
\label{respfuncspin}
\end{equation}
from which one obtains the relaxation function.

It can be easily shown \cite{Acconcia2015,acconcia2015a}  that for the family
\begin{equation}
g_\phi (t)=t/\tau+ a \ \sin\left( \kappa \pi\,t/\tau\right)\,,
\end{equation}
$\la W_\mrm{ex}\ra$ exactly vanishes. Here, $a$ and $\kappa$ are suitably chosen constants which generate zeros of $\la W_\mrm{ex}\ra$ for arbitrarily short $\tau$ \cite{Acconcia2015,acconcia2015a}.

\section{Counterdiabatic information geometry}

In a different direction Takahashi \cite{Takahashi2017} made a connection between the thermodynamic, or rather information geometry and STA. Specifically, Takahashi \cite{Takahashi2017} focused on counterdiabatic driving. In this paradigm, a system is subject to an auxiliary field, such that its wavefunction remains on the adiabatic manifold of a time-dependent Hamiltonian, $H_0(t)$. Denoting the instantaneous eigenstates of $H_0(t)$ by $\ket{n(t)}$, the total Hamiltonian can be written as \cite{Demirplak2003,Berry2009},
\begin{equation}
H(t)=H_0(t)+i\hbar \sum_n \left(\ket{\dot{n}}\bra{n}-\braket{\dot{n}}{n}\ket{n}\bra{n}\right)\,.
\end{equation}
It is easy to see that the dynamics under $H(t)$ suppresses all transitions between the $\ket{n(t)}$.

Again writing the total entropy production as relative entropy \cite{deffner2010}, and exploiting the corresponding triangle inequality Takahashi \cite{Takahashi2017} determined the irreversible entropy production solely due to the auxiliary field. Then further bounding the entropy production by the thermodynamic length \eqref{eq:qm_length}, a means to find the optimal STA protocol with only minimal dissipation becomes available. In addition, Takahashi \cite{Takahashi2017} proved
\begin{equation}
\label{eq:in_tak}
\int_0^\tau dt\,\sqrt{\e{\beta \la W_\mrm{ex}\ra}-1}\geq \sqrt{b^2\, C_\mrm{min}/2}\,\mc{L}(\rho_0,\rho_\tau)\,,
\end{equation}
where $C_\mrm{min}$ is the minimal value of $C(t)$ with
\begin{equation}
C(t)\equiv \e{-\frac{b}{3}\frac{\la \left(W-\la W\ra\right)^3\ra}{\la \left(W-\la W\ra\right)^2\ra}}\,.
\end{equation}
In conclusion, ref.~\cite{Takahashi2017} showed that  ideas from thermodynamic control not only allow to find novel paradigms for STA, but also provide means to optimize existing ones.

\section{Thermodynamic uncertainty relations}

As a final example of the ubiquity of applications of thermodynamic geometry we consider thermodynamic uncertainty relations (TUR). The TUR was originally derived for classical, stochastic systems \cite{Barato2015} and sets a bound on the magnitude of fluctuations in thermodynamic processes \cite{Gingrich2016PRL}.  For quantum systems in nonequilibrium steady states, Guarnieri \etal\, showed only recently \cite{Guarnieri2019} that the TUR is a direct consequence of the thermodynamic geometry.

In particular, Guarnieri \etal\, \cite{Guarnieri2019} again lower bounded the rate of entropy production by a quadratic form,
\begin{equation}
\label{eq:TUR_1}
\la \sigma_\mrm{ex}\ra \geq \la \mb{J}\ra^\top \Delta^{-1} \la\mb{J}\ra \,,
\end{equation}
where $\la \mb{J}\ra$ is the average current vector, and $\Delta$ is the normalized covariance matrix between different steady state currents. Realizing again that $\beta \la W_\mrm{ex}\ra=\int_0^\tau dt \la \sigma_\mrm{ex}\ra$, eq.~\eqref{eq:TUR_1} is nothing else but a thermodynamic formulation of eq.~\eqref{eq:work_qm} proposed by Scandi and Perarnau-Llobet \cite{scandi2019}.

Guarnieri \etal \cite{Guarnieri2019} then showed that eq.~\eqref{eq:TUR_1} immediately implies for a single component $J_\alpha$ of the current,
\begin{equation}
\label{eq:TUR_2}
\Delta_{J_\alpha}\,\la \sigma_\mrm{ex}\ra\geq \la J_\alpha\ra^2\,,
\end{equation}
which is the thermodynamic uncertainty relation for quantum systems in nonequilibrium steady states. 

Applying thermodynamic control to the mindset of the TUR has striking consequences. The optimal currents that minimize the entropy production, cf. eq.~\eqref{eq:TUR_1}, set the sharpest bound on the fluctuations in the current.

\section{Future directions and experiments}

In the above, we have seen that thermodynamic geometry is a versatile framework to study optimal processes with minimal dissipation, and to gain fundamental insight into the bounds and limitations set by nonequilibrium fluctuations. The open question is where we go from here. Due to the close ties of thermodynamic control and stochastic thermodynamics, it is only natural that thermodynamic control strategies have found applications in the thermodynamics of information \cite{Parrondo2015}. Paradigmatic examples are Landauer erasure \cite{proesmans2020} and optimized Szilard engines \cite{song2019}.  In a very recent experiment, Saira \etal \cite{Saira2020} applied related techniques to thermal-fluctuation-driven logical bit reset on a superconducting flux logic cell. Such experiments are essential steps towards thermodynamic computing \cite{Conte2019}, which is a promising  paradigm (and complementary to quantum computing) to overcome  the restrictions imposed by the end of Moores law.

However, we are convinced that the framework of thermodynamic control also has only barely been exploited in fully quantum dynamics. As such we remain optimistic that future work will unveil more and more applications of thermodynamic STAs.

\acknowledgments
M. V. S. Bonan\c{c}a acknowledges support from FAPESP (Funda\c{c}\~ao de Amparo \`a Pesquisa do Estado de S\~ao Paulo) (Brazil) (Grant 2019/01294-4).


\end{document}